\documentclass[twocolumn,prl,showpacs,aps]{revtex4}

\usepackage{amsmath}
\usepackage{amsbsy}
\usepackage{amsthm}
\usepackage{amssymb}
\usepackage{latexsym}

\usepackage[dvips]{color,graphicx}

\begin{document}

\title{Phase Diagram of Interacting Bosons on the  Honeycomb Lattice}
\author{Stefan Wessel}
\affiliation{Institut f\"ur Theoretische Physik III,
Universit\"at Stuttgart, Pfaffenwaldring 57, D-70550 Stuttgart, Germany.}

\begin{abstract}
We study the ground state properties of repulsively interacting bosons 
on the honeycomb lattice using large-scale quantum Monte Carlo simulations.
In the hard-core limit the half-filled system develops long ranged diagonal 
order for sufficiently  strong nearest-neighbor repulsion. This staggered
solid melts at a first order quantum phase transition into the superfluid 
phase, without the presence of any intermediate supersolid phase.
Within  the superfluid phase, both the superfluid density and the 
compressibility exhibit local minima near  particle- (hole-) density   
one quarter, while the density and the condensate fraction show inflection points in this region.
Relaxing the hard-core constraint, supersolid phases emerge for soft-core 
bosons.  The suppression of the superfluid density is found to persist for
sufficiently large, finite on-site repulsion.

\end{abstract}

\pacs{05.30.Jp, 03.75.Lm, 75.10.Jm, 75.40.Mg}
\maketitle

The properties of interacting bosons confined to periodic lattice 
structures are intensively being studied recently.
This effort is driven by both the progress in realizing such many-body systems using ultra-cold atoms 
on optical lattices~\cite{zoller,greiner,eth,bloch}, as well as the interesting phases and quantum phase transitions expected to 
emerge.

In particular, the ground state can be superfluid, Mott insulating or, in the presence of disorder, a Bose glass~\cite{fisher,batrouni91,monien}.
Recently,
based on unbiased numerical simulations,
also supersolid~\cite{square,wessel_tri,triaga,triagb,triagc,triagd,tri2,chain} and exotic valence-bond-solid phases~\cite{wessel_kago} were
confirmed to be acessible,
and unconventional quantum criticality~\cite{exotic,exotic2, tri2, wessel_kago} was discussed for such systems, in particular  on 
non-bipartite, 
geometrically frustrated two-dimensional lattices, such as the triangular or the Kagome lattice.

Concerning bosons on two-dimensional bipartite, and thus non-frustrated lattices, so far mainly
a square lattice structure has also been considered in various other studies~[19-31].
However, in light of  the recent realization of two-dimensional 
graphene, with its noticeable properties of fermions on the underlying honeycomb 
lattice~\cite{graphene1,graphene2,graphene3,graphene4}, the question  arrises, how interacting bosons 
behave on the honeycomb lattice. While both the honeycomb and the square lattice are bipartite, quantum flucuations are expected to be more relevant on the honeycomb lattice, due to its lowest possible coordination in two dimensions.

Motivated by these considerations, we  analyze in this Letter the ground state  properties of the 
extended 
Bose-Hubbard model
\begin{eqnarray}
H&=&-t \sum_{\langle i,j \rangle} \left( b^\dagger_i b_j + b^\dagger_j b_i \right)
  +\frac{U}{2} \sum_i  n_i (n_i-1)\nonumber\\
  &&+V \sum_{\langle i,j \rangle} n_i n_j
  -\mu \sum_i n_i
\end{eqnarray}
on the honeycomb lattice. 
Here, $b^{\dagger}_i$ ($b_i$) denote bosonic 
creation (annihilation) operators for bosons on lattice site $i$, $t$ is the 
nearest-neighbor hopping amplitude,
$U$ 
 an onsite repulsion, $V$ a nearest-neighbor repulsion,  and $\mu$ the chemical potential in 
the grand-canonical ensemble, which controls the filling of the lattice.

The honeycomb lattice is  bipartite,  
with a two-site unit cell and a uniform coordination number $z=3$; the 
unit cell contains one site from each sublattice.
In the following, 
we employ the stochastic series expansion~\cite{sse} quantum Monte Carlo (QMC) method with directed loop updates~\cite{directed,directed2}, and study the Hamiltonian 
in Eq.~(1) on 
finite lattices with $N=2L^2$ sites for $L$ up to 32,  at temperatures 
sufficiently low in order to resolve ground state properties of these 
finite systems~\cite{square}. In contrast to geometrically frustrated lattices, the sign  of $t$ in Eq.~(1) is irrelevant on the bipartite honeycomb lattice, in that also $t<0$ would cause no QMC sign problem. 

\begin{figure}[t]
\includegraphics[width=.45\textwidth]{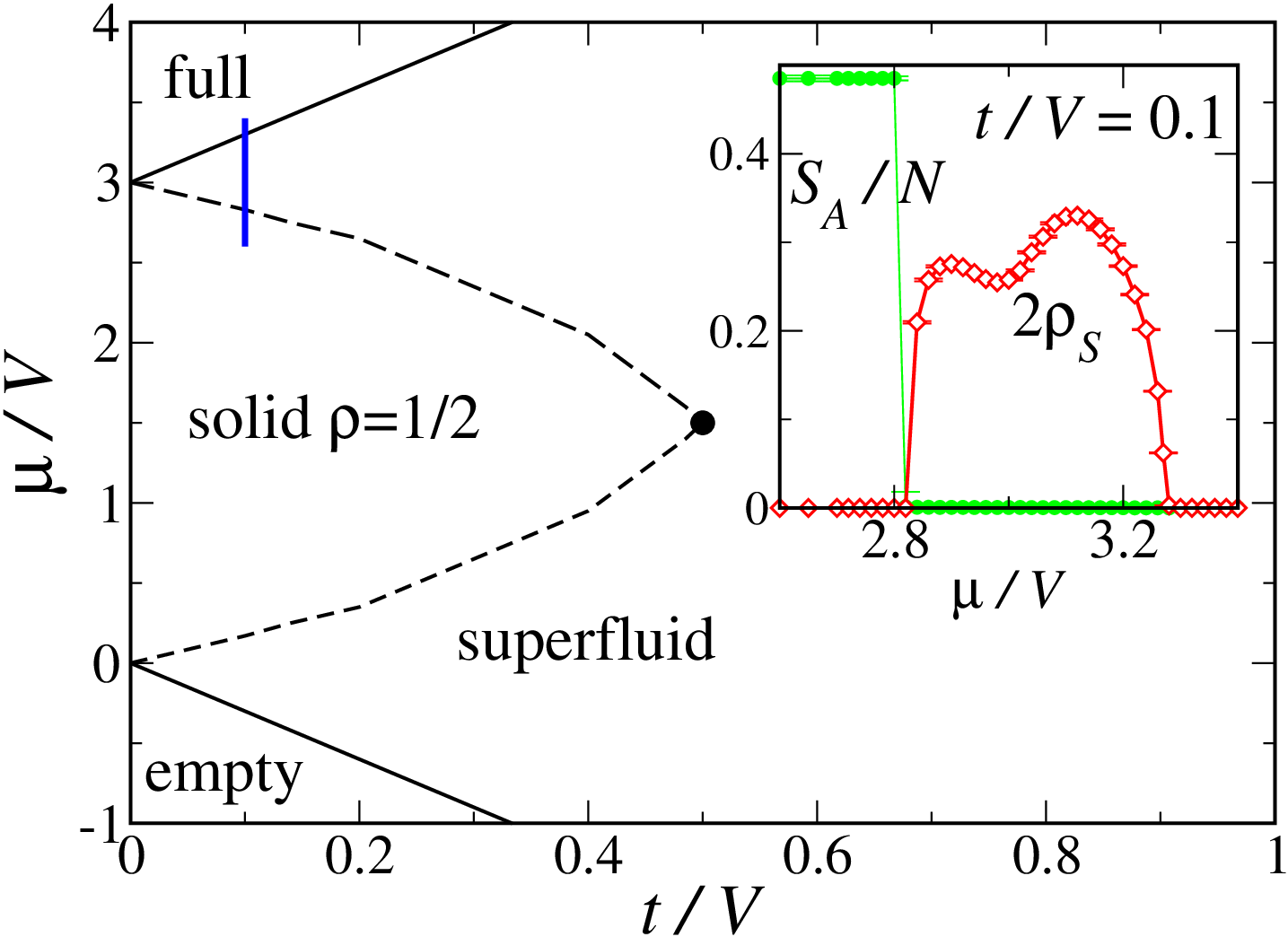}
\caption{(Color online) Ground state phase diagram of  bosons on the honeycomb lattice in the hard-core limit. 
The black dot locates the Heisenberg point, where the model has an enhanced $SU(2)$ symmetry.
In the inset the superfluid density $\rho_S$ and the  structure factor $S_A$ are shown along 
$t/V=0.1$ (indicated by the vertical bar in the main part of the figure).}
\end{figure}

We first consider the hard-core limit, $U/t\rightarrow \infty$, of Eq.~(1), for which
the  bosonic model can be mapped onto the spin-$1/2$ XXZ model~\cite{mapping}.
This allows
for an interpretation of
our results in terms of both bosons
and quantum spins.
The ground state phase diagram of the grand-canonical ensemble in the hard-core limit
is shown 
in Fig.~1. 
Considering the one-particle and one-hole problem, one finds that
for $\mu<-zt$ the system is empty, and for 
$\mu/V>z(1+t/V)$ 
it is fully occupied by one boson per lattice site (density $\rho=1$). At large values of
$t/V>0.5$, 
the bosons form a superfluid (SF) with off-diagonal long-range order (ODLRO), 
characterized by a finite superfluid 
density, which in the QMC simulations is obtained as 
$\rho_s=\langle W^2\rangle/(2\beta t)$
from the 
particle winding number fluctuations $\langle W^2\rangle$, where $\beta$ denotes the inverse temperature~\cite{windingnumber}.
Furthermore, for sufficiently small values of 
$t/V$, the system exhibits a solid phase of density $\rho=1/2$. 
Its largest extend is set by the 
Heisenberg point $(t/V,\mu/V)=(1/2,3/2)$, where the Hamiltonian in Eq.~(1) has an enhanced $SU(2)$ 
symmetry.
In the perfect $\rho=1/2$ solid at $t=0$, one out of the two sublattices on the honeycomb lattice is occupied, 
the other being empty (so that reflection symmetry, but not lattice translational symmetry is broken). 
The corresponding structure factor 
for this alternating (staggered) diagonal long-range order (DLRO) is thus
given by 
\begin{equation}
S_A=\frac{1}{N} \sum_{i,j} \epsilon_i \epsilon_j 
\langle n_i n_j\rangle ,
\end{equation}
where $\epsilon_i=\pm 1$ for $i$ on sublattice $A (B)$,
in terms of the  density-density correlations.
\begin{figure}[t]
\includegraphics[width=.45\textwidth]{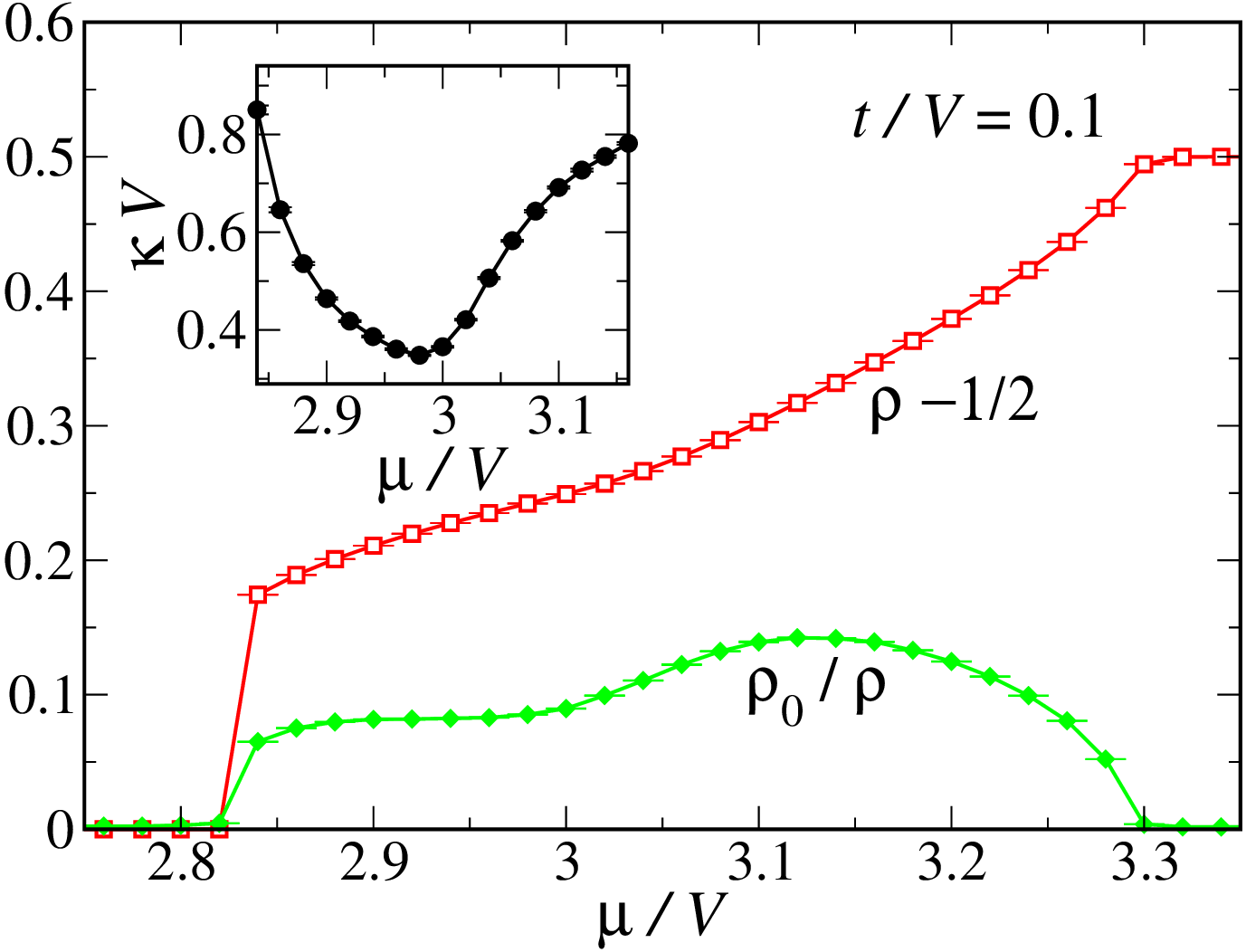}
\caption{(Color online) Density $\rho$ (with respect to half-filling) and condensate fraction $\rho_0/\rho$ at $t/V=0.1$ as functions of $\mu/V$ for  bosons on the honeycomb lattice in the hard-core limit. 
The inset shows the compressibility $\kappa$ which exhibits a local minimum in the superfluid regime.}
\end{figure}
The inset of Fig.~1 shows the evolution of both $\rho_S$ and the order parameter $S_A/N$ along 
a cut through the phase diagram at $t/V=0.1$ (the vertical bar in 
the main 
part of Fig.~1), and corresponds to their thermodynamic limit behavior. 
Both quantities clearly exhibit the  first-order nature of the 
solid-SF quantum phase transition, and the absence of any intermediate 
supersolid  (SS)
phase with simultaneous DLRO and ODLRO (only at the Heisenberg point  may DLRO and ODLRO co-exist, due to the 
$SU(2)$ symmetry).
In the hard-core limit, the phase diagram of Eq.~(1) is
thus similar to the 
one of hard-core bosons on the square lattice~\cite{square}, irrespectively  a rescaled  $\mu$-axis, due to the different coordination number ($z=4$ on the square lattice). In both cases, domain wall proliferation at the quantum melting transition of the half-filled solid  renders a SS 
unstable towards phase separation in the hard-core limit of Eq.~(1)~\cite{square}. Note, that also through the Heisenberg point the quantum melting of the $\rho=1/2$ solid is discontinuous~\cite{herbert,kuklov, four-site}, despite  attempts to extract critical behavior from  finite size scaling plots of $\rho_S$~\cite{herbert, gan}. In fact, the extracted effective dynamical critical exponent $z$
decreases to zero, taking sufficiently large system sizes~\cite{four-site},  concistent with the trend observed by comparing the values obtained thus far~\cite{herbert, gan}.\\

In contrast to the case of the square lattice, we observe on the honeycomb lattice a pronounced 
dip of
$\rho_S$  inside the SF phase (c.f. the inset of Fig.~1), with a local minimum at a 
filling $\rho_D$ close to  $3/4$, corresponding to a hole density of $1/4$ ($\rho_D\approx 0.744$ at $(\mu/V)_D\approx 2.97$ for $t/V=0.1$). Due to particle-hole symmetry in the hard-core limit, this suppression of $\rho_S$ also appears  near a particle filling of $\rho=1/4$. 
Fig.~3 shows a similar dip in $\rho_S$ for $t/V=0.2$, where we again find $\rho_D$  close to $3/4$.
\begin{figure}[t]
\includegraphics[width=.48\textwidth]{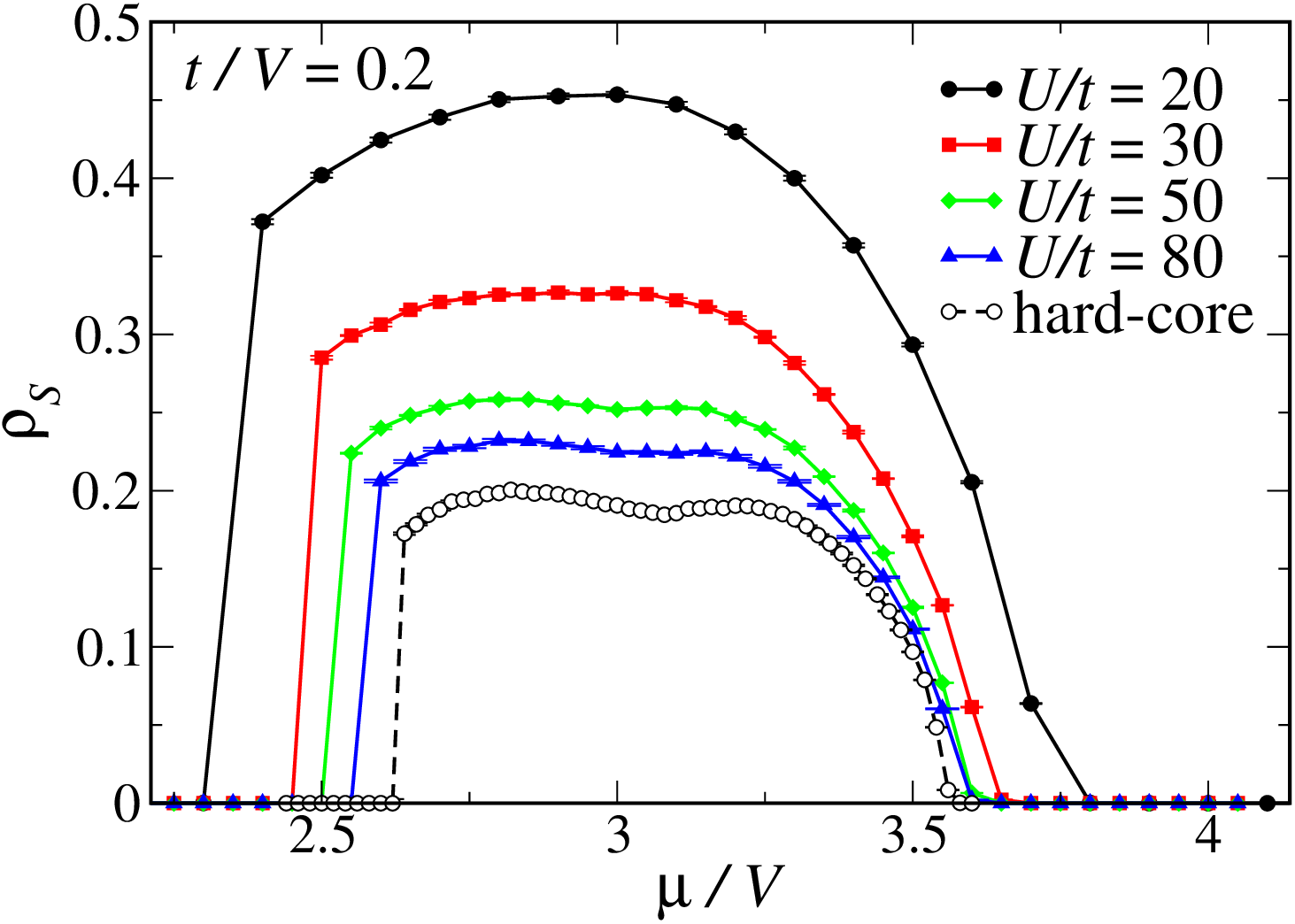}
\caption{(Color online) Superfluid density of bosons on the honeycomb lattice at $t/V=0.2$ as functions of $\mu/V$ for various values of $U/t$, and in the hard-core limit.}
\end{figure}
This indicates the presence of 
geometric hindrance in the superfluid flow on the honeycomb lattice, well inside the SF region.  Indeed, 
also the 
compressibility $\kappa=\partial \rho / \partial \mu$ exhibits a minimum 
at $(\mu/V)_D$, c.f. the inset of Fig.~2. In contrast, as seen in Fig.~2, both the density $\rho$ and the condensate 
fraction, $\rho_0/\rho$, where $\rho_0=\langle b^\dagger({\mathbf{k}=0})b({\mathbf{k}=0})\rangle$, 
increase in that region upon increasing $\mu$, and show an inflection point at $(\mu/V)_D$. 
We verified that near the dip the system does {\it not} exhibit any 
incommensurate long-range order neither in the density-density  
nor in 
bond-bond correlation functions~\cite{wessel_kago}.\\

\begin{figure}[t]
\includegraphics[width=.48\textwidth]{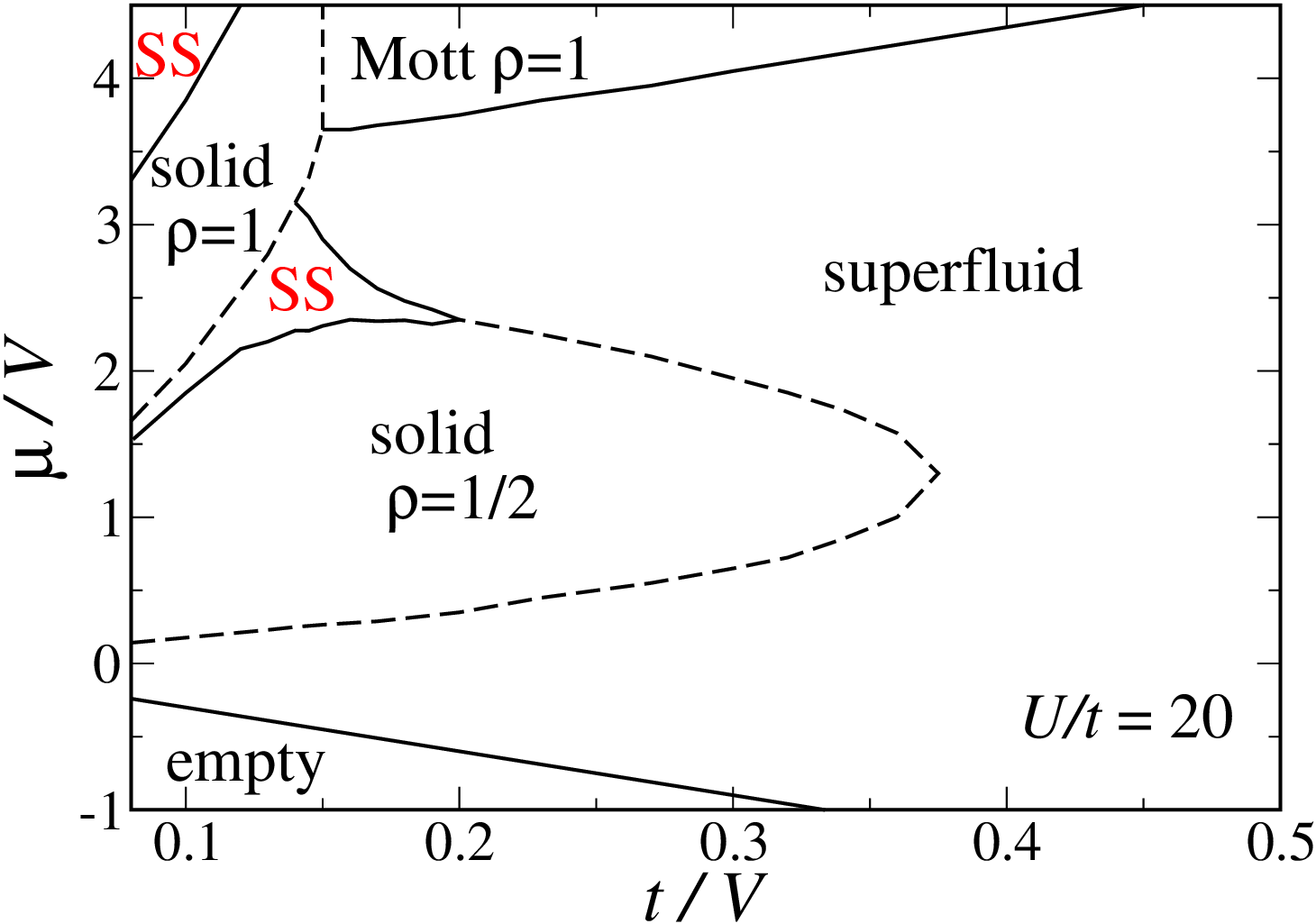}
\caption{(Color online) Ground state phase diagram of  bosons on the honeycomb lattice for finite $U/t=20$, with supersolid (SS) phases emerging upon doping the solid phases. 
Solid (dashed) lines denote continuous (first-order) quantum phase transitions.}
\end{figure}
Extending the analysis beyond the hard-core limit, we next consider  soft-core bosons on the honeycomb lattice, and  assess the presence of SS phases in this model. Fig.~4 shows the ground state phase diagram for $U/t=20$ inside the regime where $\rho \leq 1$. The phase diagram exhibits  several additional phases; in particular, a SS phase emerges upon doping the $\rho=1/2$ solid with additional bosons. The simultaneous presence of DLRO and ODLRO within the SS regime for $t/V=0.14$ and $t/V=0.16$ can be seen from Fig.~5, by  finite values of both $S_A/N$ and $\rho_S$ inside a range of $\mu/V$ values.
For $(zV-U)\sim t$, the additional particles form  a superfluid atop the $\rho=1/2$ solid background, due to a kinetic energy gain which prevents the domain wall proliferation~\cite{square}. Hole doping of the $\rho=1/2$ solid however does not result in a similar kinetic energy gain, and thus no SS state emerges for $\rho<1/2$.

The nature of the incompressible state with $\rho=1$ depends on the ratio $zV/U$~\cite{square}:
In the soft-core case, taking $U/t=20$, we find  for $t/V<0.15$ (i.e., $zV/U>1$), that the system forms a $\rho=1$ solid  with a finite value of $S_A/N$ in the thermodynamic limit. This  $\rho=1$ solid corresponds to each site of one out of the two sublattices being occupied by two bosons, the other sublattice being empty. In contrast, for $t/V>0.15$ (i.e., $zV/U<1$), the system is a uniform $\rho=1$ Mott insulator,
with $S_A/N=0$. This different behavior at $\rho=1$ is seen for $t/V=0.14$ and $t/V=0.16$ in Fig.~5, and results from the competition between onsite- and nearest-neighbor repulsion terms in Eq.~(1)~\cite{square}.
Fig.~5 furthermore indicates, that the transition from the $\rho<1$ SS to the $\rho=1$ solid is strongly first-order, whereas the SS-SF transition and the SF-Mott transition are continuous, as  expected also from kinetic energy considerations~\cite{square}.
As indicated in Fig.~4, a  further SS  phase, with $\rho>1$, emerges out of the $\rho=1$ solid phase upon further increasing $\mu$. Details of the phase diagram for fillings $\rho>1$ will be presented elsewhere~\cite{wessel_more}.

\begin{figure}[t]
\includegraphics[width=.45\textwidth]{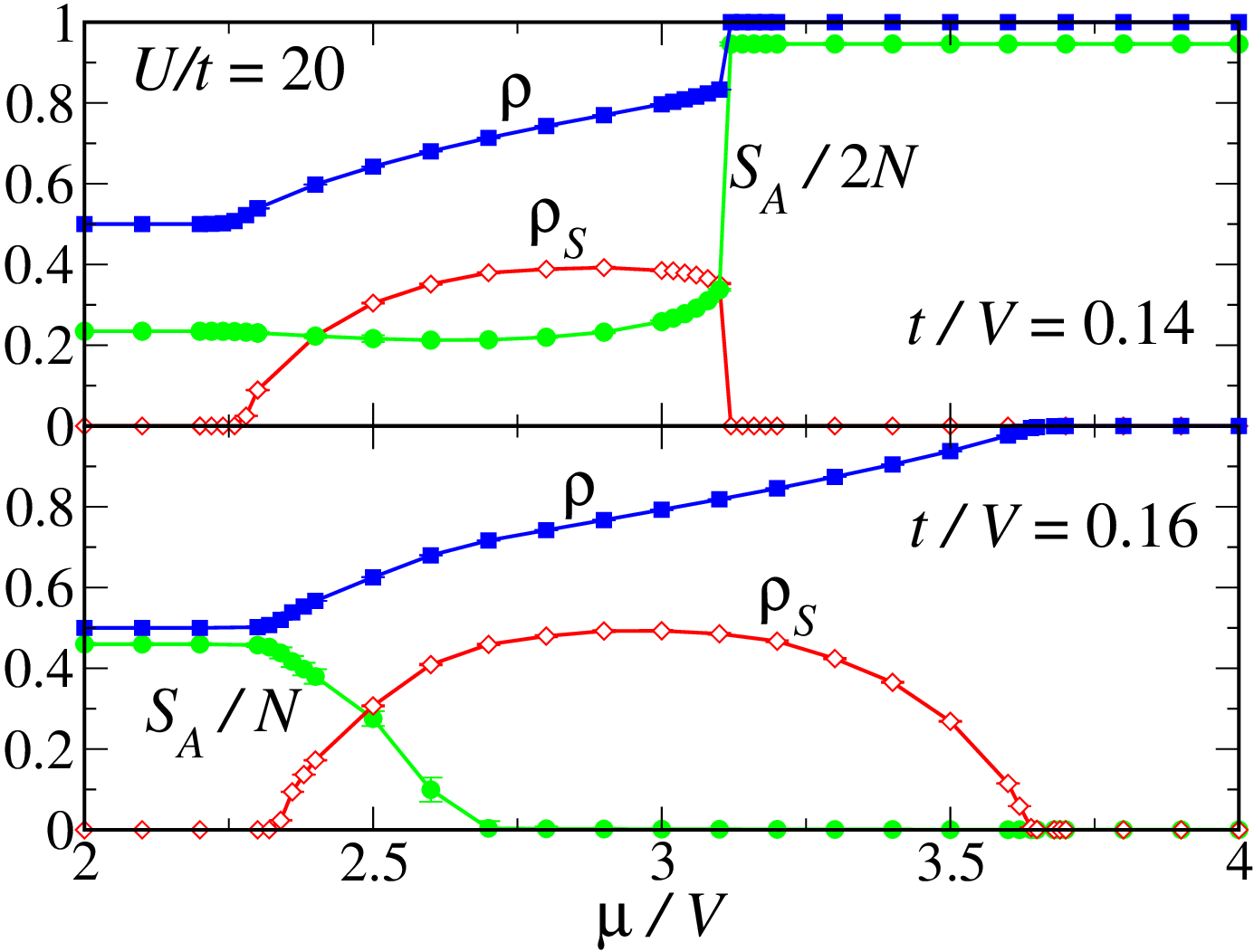}
\caption{(Color online) Density $\rho$, superfluid density $\rho_s$, and structure factor $S_A$ as functions of $\mu/V$ for bosons on the honeycomb lattice for $U/t=20$ along $t/V=0.14$ (upper panel) and $t/V=0.16$ (lower panel).}
\end{figure}
For $U/t=20$, the superfluid density curves in Fig.~5 do not exhibit  local minima such as the pronounced dip in $\rho_S$ observed in the hard-core limit (in the inset of Fig.~1). In order to assess, whether this suppression is restricted to the hard-core limit of Eq.~(1), we performed simulations also for increasing values of $U/t=30, 50$, and 80. In Fig.~3, the resulting values of $\rho_S$ are shown as functions of $\mu/V$ for a common value of $t/V=0.2$, and compared to the hard-core limit. We find that for $U/t=30$ the superfluid density is almost flat near $\mu/V\approx 2.9$, and  for sufficiently large onsite repulsion $U$, it  indeed develops a shallow local minimum, similar to the dip in the hard-core limit. The reduced superfluidity thus relates to the interplay of a
large energy penalty for double occupation, and the low connectivity on the honeycomb lattice, compared to the square lattice. In fact, we find that on (inhomogeneous) two-dimensional lattices with an even lower average coordination number $\bar{z}<3$, the superfluid density gets reduced  to zero, and incompressible phases with fractional fillings $\rho=1/4$ and $3/4$ emerge~\cite{wessel_more}.\\

In conclusion, we studied the ground state properties of bosons on the honeycomb lattice, as described by the extended Bose-Hubbard model. We found that the phases and the structure of the phase diagram correspond to those of the same model on the square
lattice, both in the hard-core limit and for finite on-site repulsion $U$. 
This is anticipated given the bipartiteness of both lattice structures.
In particular, a staggered solid phase exists at half-filling for sufficiently strong nearest-neighbor repulsion $V$. Doping this solid with additional bosons leads to a supersolid phase only for finite values of $U$, whereas in the hard-core limit and for hole-doping the system phase separates. The incompressible state at  filling one is a Mott insulator for dominant $U>zV$, and a staggered solid  for $U<zV$. 
Furthermore, we found that for sufficiently large values of $U$, both the superfluid density and the compressibility are suppressed  inside the superfluid phase at particle fillings of about one quarter and three quarters. As the condensate density does not show a similar reduction, bosons on the honeycomb lattice thus feature regions, where an {\it increase} of the condensate in the bosonic system contrasts to a {\it decrease} in its superfluid response.\\

We would like to thank F. Alet, R. Moessner, and A. Muramatsu for helpful discussions, and NIC J\"ulich and HRLS Stuttgart for allocation of CPU time.\\

{\it Note added}.-
After finishing the numerical calculations, we became aware of a parallel work~\cite{gan}, where results partially similar to our findings are presented.

\end{document}